\documentclass[conference,onecolumn]{IEEEtran}
\usepackage[dvips,final]{graphicx}
\usepackage{epsfig,amssymb,color,amsbsy,amssymb,amsmath}
\newcommand{\beq}{\begin{equation}}
\newcommand{\eeq}{\end{equation}}
\newcommand{\beqn}{\begin{eqnarray}}
\newcommand{\eeqn}{\end{eqnarray}}

\long\def\symbolfootnote[#1]#2{\begingroup%
\def\thefootnote{\fnsymbol{footnote}}\footnote[#1]{#2}\endgroup}

\title{Subspace Techniques for Radio-Astronomical Data Enhancement}
\author{Sarod Yatawatta$^1$\\ Kapteyn Institute, University of Groningen,\\ and ASTRON,\\ The Netherlands.\\
\thanks{This work
was supported by LOFAR and SNN. LOFAR is being funded by the European Union, European Regional Development Fund, and by ``Samenwerkingsverband Noord-Nederland'', EZ/KOMPAS.}
 yatawatta@astron.nl}
\begin{document}
\maketitle

\symbolfootnote[0]{$^1$This work was first presented in the SKA calibration and imaging workshop, Cape Town, SA, December 2006 and in URSI General Assembly, Chicago, US, August 2008}

\begin{abstract}
Radio astronomical observations have very poor signal to noise ratios, unlike in other disciplines. On the other hand, it is possible to observe the object of interest for long time intervals as well as using a wider bandwidth. Traditionally, by averaging  in time and in frequency, it has been possible to improve the signal to noise ratio of astronomical observations to improve the dynamic range. This is possible due to the inherent assumption that the object of interest in the sky is invariant over time and the frequency range of observation. However, in reality this assumption does not hold, due to intrinsic variation of the sky as well as due to errors generated by the instrument. In this paper, we shall discuss  an alternative to averaging of images, without ignoring subtle changes in the observed data over time and frequency, using subspace decomposition. By separation of data to signal and noise subspaces, not only would this improve the quality of the data, but also enable us to detect faint artifacts due to calibration errors, interference etc.
\end{abstract}

\section{Introduction}
Astronomical observations have very weak signal to noise ratios (SNR). In radio astronomy (interferometry), this weak SNR is improved mainly by  prolonging the observation time. This not only improves the SNR, but also sampling of the $uv$ plane. A typical observation has a bandwidth of a few MHz, divided into smaller narrowband channels. Normally the bandwidth of a single channel is about a few kHz. After interference mitigation, calibration and imaging (also deconvolution) each of these narrow band channels, it is normal procedure to average them, on the image plane. The resultant averaged image has an improved SNR compared to images made by each of the narrowband channels, assuming all systematic errors have been removed. This is the typical process by which current radio telescopes like the Westerbork Synthesis Radio Telescope (WSRT), is able to achieve dynamic  ranges in the order of 1,000,000 to 1.

However, there is a significant drawback in this averaging procedure, because it is based on a erroneous assumption: that the source as well as the effects of the instrument remains invariant in all averaged data. Only the noise varies and averaging significantly decreases the noise power. The invariance assumption is accurate to some extent, but closer scrutiny reveals that there is significant amount of variation:
\begin{itemize}
\item There is intrinsic variation of the  sky (or the sources) with frequency (non zero spectral indices).
\item The $uv$ points scale with frequency, thus the point spread function (PSF) change in the images.
\item Transient sources as well as radio frequency interference (RFI) create temporal variation.
\item Calibration errors, small as they might be, create subtle variations in time as well as in frequency.
\end{itemize}

Indeed, by averaging, both noise and the aforementioned effects get suppressed.
However, it is to our advantage that we detect such variations in our images. For instance, we could detect trace amounts of RFI and calibration errors etc., that appear only in certain time or frequency ranges. This could also enable us to discard certain parts of the data, where such effects are dominant, and thus improve the quality of our averaged, final image. 

Subspace techniques have previously being used in image compression \cite{Andrews76, Yang95}, as well as image comparison. However, to the best of our knowledge, this has not been used in (radio) astronomical image processing.

The rest of the paper is organized as follows: In the next section, we give a formal representation of images and subspace decomposition. Next, we investigate various uses of the image subspace decomposition. Later, we consider possible extensions of the proposed method, by taking into account the data weights as well as the affect of PSF. Finally, we give our conclusions.

Notation: All bold lowercase letters represent a column vector, all bold uppercase letters represent a matrix.

\section{Mathematical Preliminaries}
Let us represent an arbitrary image, i.e., a set of pixels with arbitrary arrangement in two dimensions, as a vector $\bf b$. Let the number of chosen pixels be $N$, so the dimension of vector $\bf b$ is $N \times 1$. Instead of one such image, we have a set of images, of the same part of the sky. Let us denote this set by $\mathcal I$. Let us assume we have $M$ such images, which could differ in terms of any of the following:
\begin{itemize}
\item Observing frequency, one image per each channel.
\item Observing time, different observation dates.
\item Observing instrument, using different radio telescopes, i.e. WSRT or the Very Large Array (VLA).
\item Integration time, using snapshots
\end{itemize}

In other words, the images in the set $\mathcal I$ have one or more forms of diversity. Let ${\bf b}_j$ represent the $j$-th such image from $\mathcal I$. We assume all the images in $\mathcal I$ could be represented by an arbitrary orthonormal basis $\bf V$ ($N \times N$ matrix), as in (\ref{basis}). Note that $\bf V$ is used to represent the components visible in the image, i.e., the signal, not the noise.
\beq \label{basis}
{\bf b}_{j} = {\bf V} {\bf a}_j + {\bf n},\ \ j\in [1\ldots M]
\eeq

In (\ref{basis}), ${\bf a}_j$ represent the components of the basis $\bf V$ used to represent the $j$-th image. The noise, which is assumed to be independent in each image in $\mathcal I$ is represented by $\bf n$ ($N \times 1$ vector). Moreover, the signal or the artifacts are represented by ${\bf V} {\bf a}_j$ in (\ref{basis}).

We make the following assumptions:
\begin{itemize}
\item The variations in all the images in $\mathcal I$ is small enough to be represented by at most $N$ orthonormal vectors.
\item The noise in each image a stationary random process with autocorrelation $\sigma^2 {\bf I}$, and is uncorrelated between pixels as well as with the signal.
\item The noise power is much lower compared to the signal or artifacts.
\end{itemize}

Let us consider the autocorrelation $\bf R$ of the images, given by (\ref{auto}).
\beq \label{auto} 
{\bf R}\buildrel \triangle \over =E\{{\bf b}{\bf b}^{T}\}
\eeq
Using $M$ images, the estimate of  the autocorrelation $\hat{\bf R}$ is given by (\ref{est}).
\beq \label{est}
\hat{\bf R} = \frac{1}{M} \sum_{j=1}^{M} {\bf b}_j{\bf b}_j^{T}
\eeq

We exploit the assumption that noise is uncorrelated with signal, i.e.,
 $E\{{\bf n}{\bf a}^{T}\}=E\{{\bf a}{\bf n}^{T}\}={\bf 0}$.
Because ${\bf V}{\bf V}^{T}={\bf I}$, and   we get
\beq \label{fun}
{\bf R}={\bf V} E\{{\bf a}{\bf a}^T\}{\bf V}^{T}+\sigma^2{\bf I}
={\bf V}\left( E\{{\bf a}{\bf a}^T\}+\sigma^2{\bf I}\right){\bf V}^{T}.
\eeq

From (\ref{fun}), we see that the column space of ${\bf R}$ is a subset of the column space ${\bf V}$. 
\beq
col({\bf R}) \subseteq col({\bf V})
\eeq
Using the Singular Value Decomposition (SVD) we can find an equivalent column space
 ${\bf R}={\bf U}{\bf \Sigma}{\bf U}^{T}$  as in (\ref{col}). 
\beq \label{col}
col({\bf U}) \subseteq col({\bf V})
\eeq

The singular values in ${\bf \Sigma}$ will characterize the signal and noise subspace eigenmodes. Because the noise has much lower power compared with the signal, the eigenmodes corresponding to the dominant singular values in ${\bf \Sigma}$ will represent the subspace of $\bf V$.

Hence, we can divide eigenvectors as noise or signal according to their singular values (\ref{signoise}), where ${\bf U}_s$ and ${\bf U}_n$ represent the signal and noise subspaces, respectively.
\beq \label{signoise}
 {\bf U} =\left[ \ {\bf U}_s \ \ \vdots \ \ {\bf U}_n \ \right]  
\eeq

Given an image ${\bf b}_j$, we can separate the signal ${\bf b}_{sj}$ and noise ${\bf b}_{nj}$ as in (\ref{sn}).
\beq \label{sn}
{\bf b}_{sj}= {\bf U}_s {\bf U}_s^{T} {\bf b}_j\ \ \
{\bf b}_{nj}= ({\bf I}-{\bf U}_s {\bf U}_s^{T}){\bf b}_j
\eeq

Note in practice, in (\ref{est}), we normally have fewer images than the image size in pixels, i.e., $N \gg M$. Hence, we would only have $M$ non zero singular values in ${\bf \Sigma}$. However, by our assumption (small enough variation), it should be possible to represent the signal space using fewer eigenmodes than $M$.
\section{Applications}
The aforementioned technique has been widely used in image compression and comparison in other disciplines. However, we could find novel uses of this technique in radio astronomy.
\begin{itemize}
\item Enhancement of signal to noise ratio: For each image ${\bf b}_j$ in the set $\mathcal I$, we get the enhanced version ${\bf b}_{sj}$ in (\ref{sn}). These images could be used for further processing, like estimation of spectral indices of fainter sources.
\item Automatic image classification: A typical observation (say in LOFAR) consists of a few thousand channels, with an image being made for each one. It is assumed the majority of these images are good quality, with proper calibration and imaging. However, a few of these images would still have faint RFI or calibration errors. In order to automatically determine image quality, we could use the power of the noise component, i.e., the norm of ${\bf b}_{nj}$  in (\ref{sn}). For a poor quality image, this should be much higher compared to a good quality image.
\item Continuum subtraction, spectral lines: Let us assume that the $j$-th image contains a weak spectral line, hidden under a strong continuum signal that is common to all the images in the set $\mathcal I$. Just by looking at this image ${\bf b}_j$, it might not be possible to observe this spectral line because of the continuum signal. However, by subtracting this, in ${\bf b}_{nj}$, we should be able to observe this.
\item Detection of faint, narrowband RFI: The same technique applied to spectral lines could be applied to detect faint, narrowband RFI. It is assumed that the RFI affects only a few images in the set $\mathcal I$.
\end{itemize}

An implementation of this method which is capable of decomposing FITS files can be downloaded from \cite{url}.
\section{Example}
\begin{figure}[htb]
\begin{minipage}[b]{0.48\linewidth}
\centering
 \centerline{\epsfig{figure=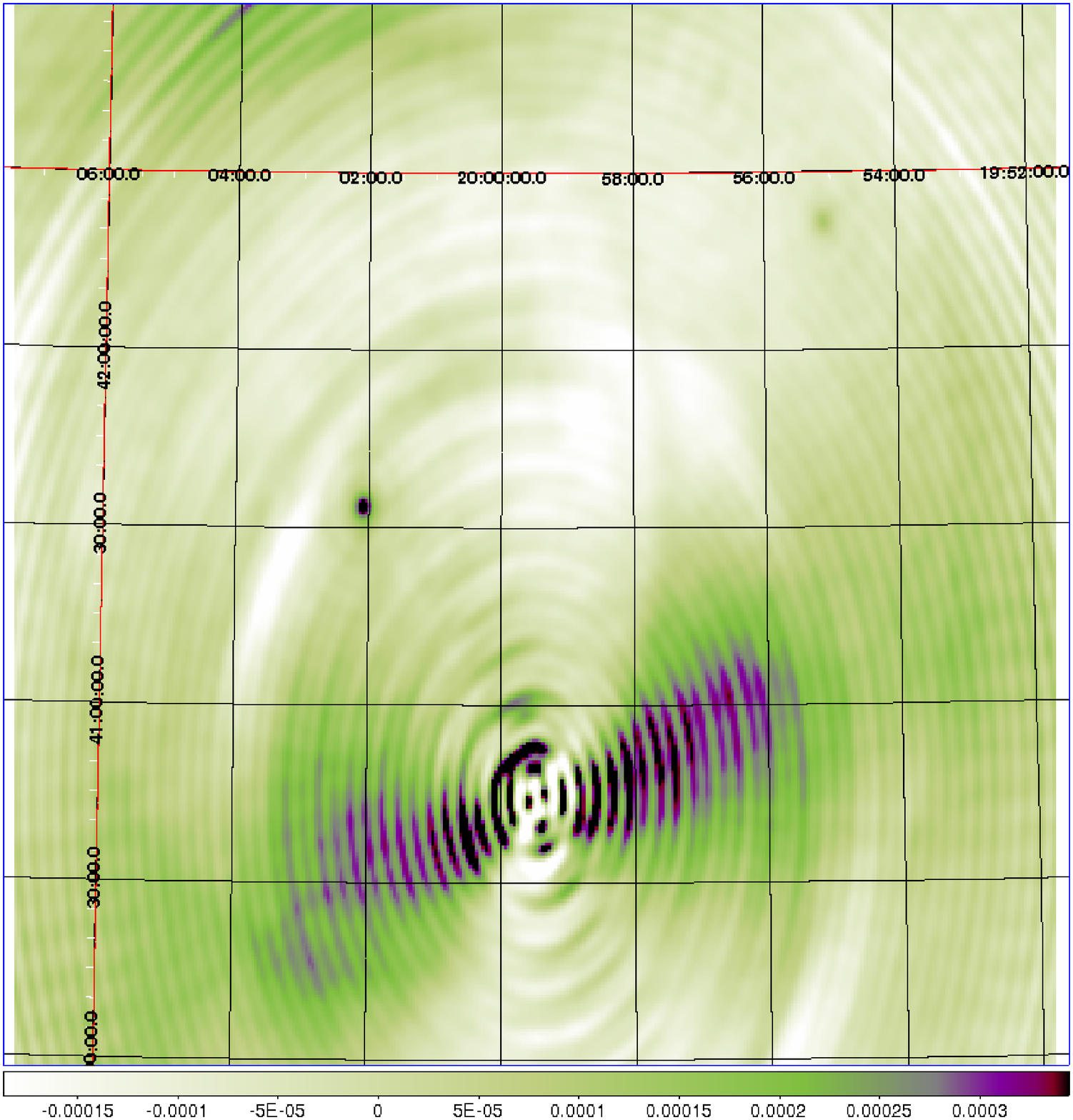,width=6.0cm}}
\vspace{0.5cm} \centerline{(a)}\smallskip
\end{minipage}
\begin{minipage}[b]{0.48\linewidth}
\centering
 \centerline{\epsfig{figure=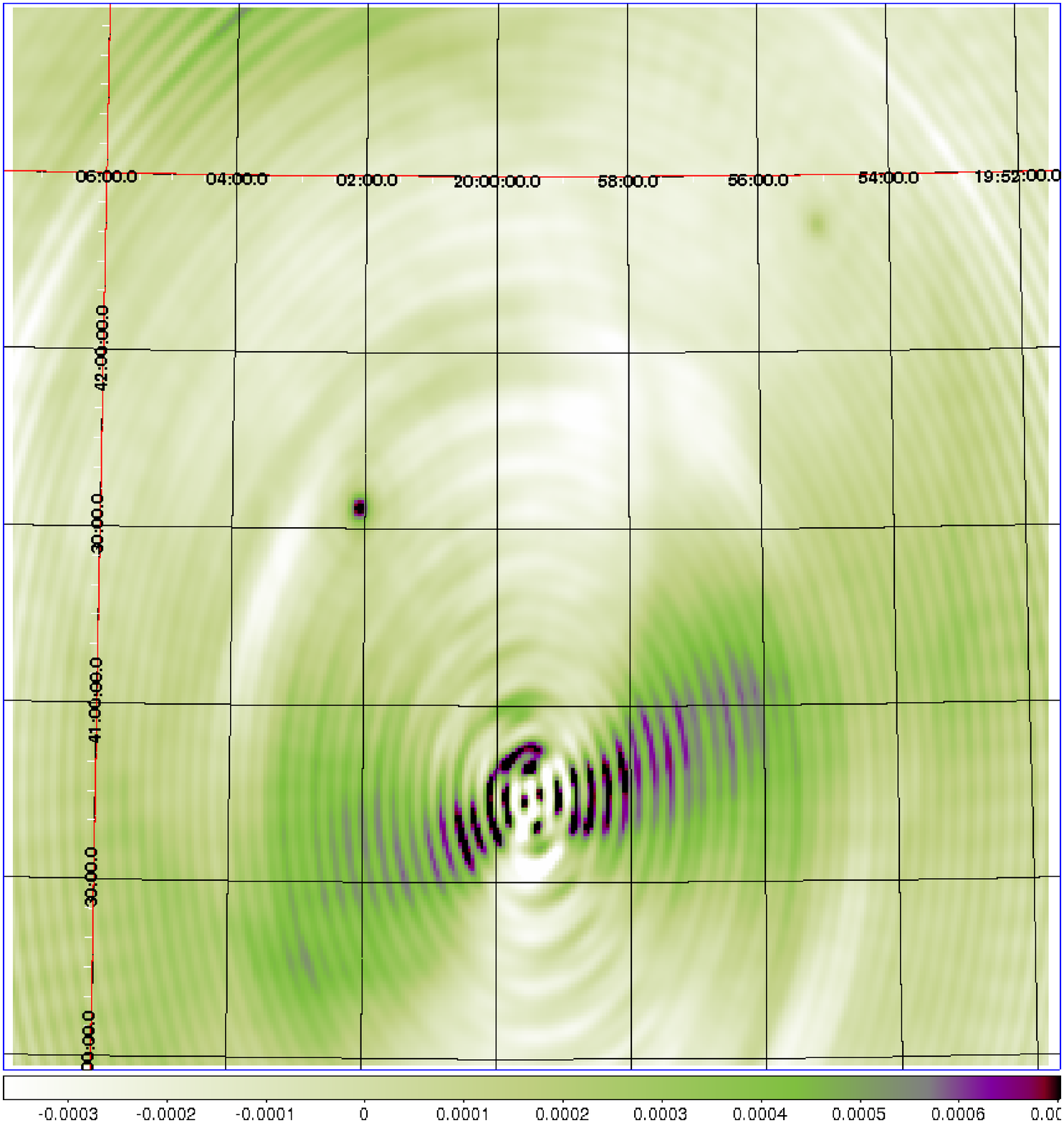,width=6.0cm}}
\vspace{0.5cm} \centerline{(b)}\smallskip
\end{minipage}
\\
\begin{minipage}[b]{0.48\linewidth}
\centering
 \centerline{\epsfig{figure=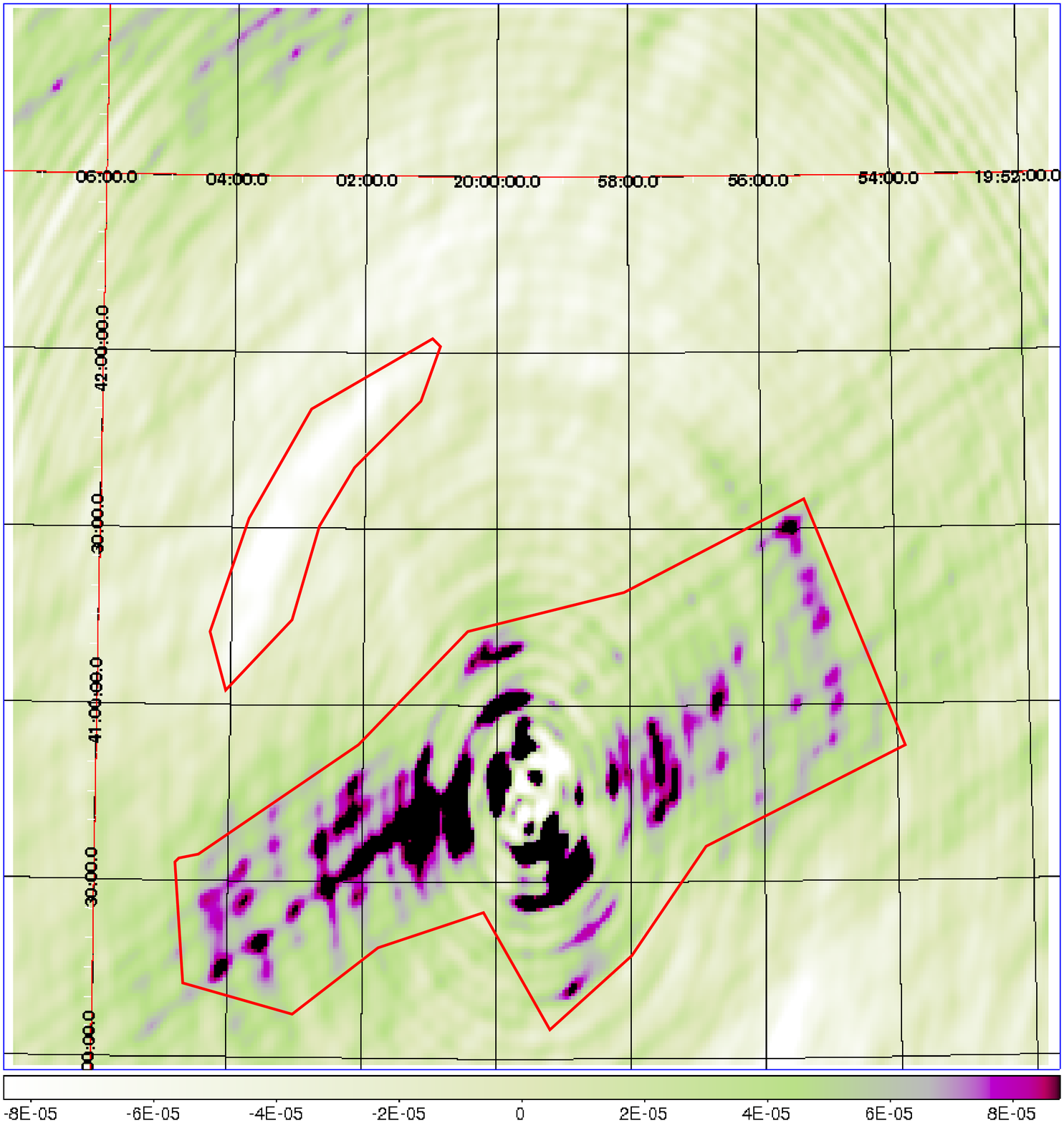,width=6.0cm}}
\vspace{0.5cm} \centerline{(c)}\smallskip
\end{minipage}
\begin{minipage}[b]{0.48\linewidth}
\centering
 \centerline{\epsfig{figure=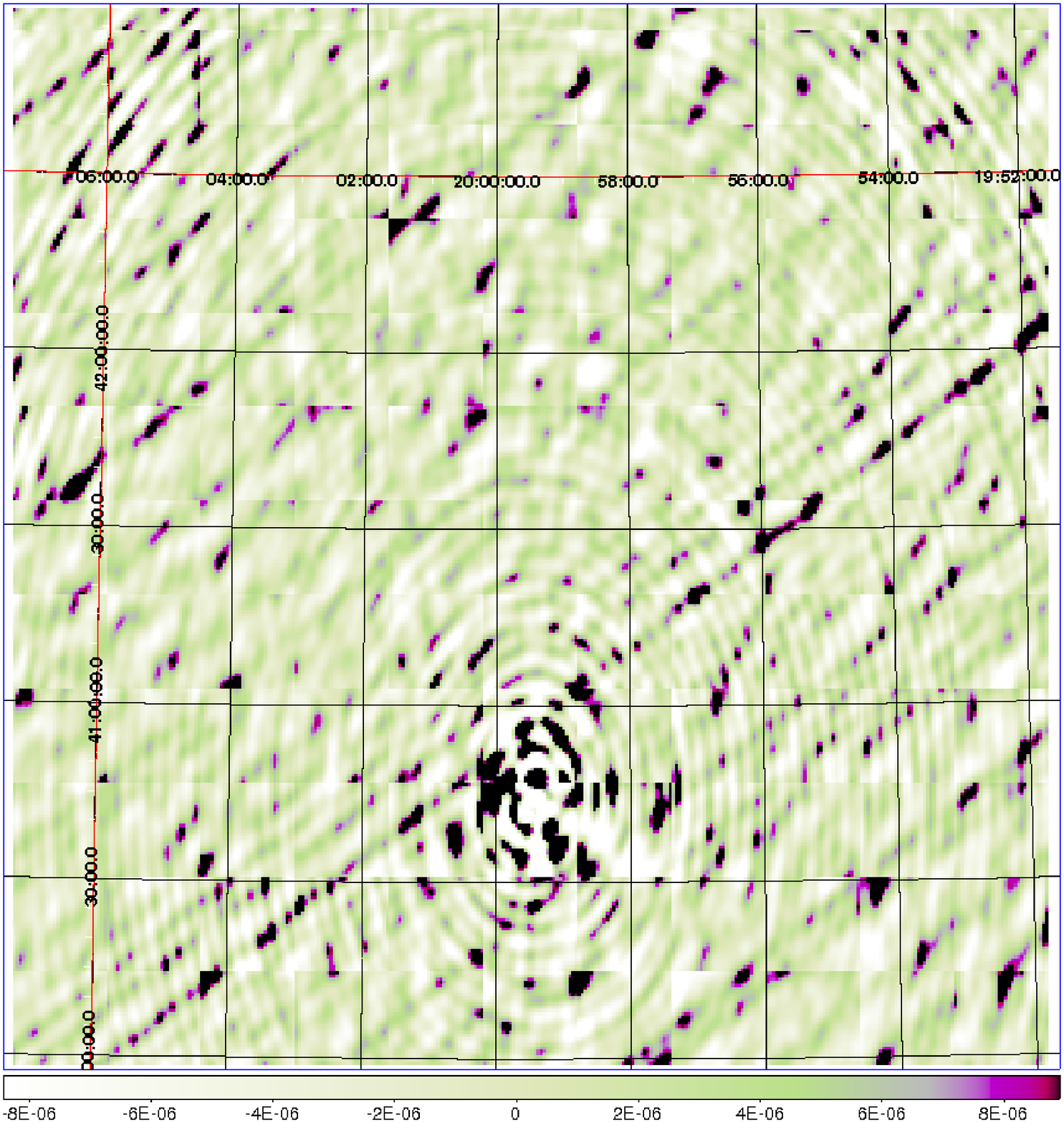,width=6.0cm}}
\vspace{0.5cm} \centerline{(d)}\smallskip
\end{minipage}
\caption{Partially deconvolved residual image around Cyg A using WSRT data: \{(a)\}  residual image made by a single channel of data, \{(b)\} mean image using 32 channels, \{(c)\} difference image between (a) and (b), \{(d)\} residual remaining after removal of signal subspace from the image \{(a)\}. It is clearly evident that the image (a) contains faint RFI, seen on (d) as diagonal lines in black. These appear also on the difference image (c) but it is hard to distinguish them because of the artifacts remaining. The artifacts on (c) are outlined using the red polygons. By using the proposed technique, we are able to remove most of these artifacts. The tiled structure on (d) is due to the divide-and-conquer approach in applying this technique.} \label{real}
\end{figure}

In this example, we have considered a WSRT observation with faint narrowband RFI, which has not been detected by the flagging algorithm. We have about 32 channels of data, each with a bandwidth of 100 kHz centered around 140 MHz. We have applied the proposed technique on this dataset. The results are given on Fig. \ref{real}.

In order to reduce the computational cost (for an image with $N$ pixels, the cost of computation of the SVD is $O(N^3)$), we have used a divide and conquer approach. Given a set of images with a large number of pixels, we first divide them into smaller sub-images, using an arbitrary boundary. For each set of these sub images (one taken from each large image), we apply the proposed technique. This saves the computational cost significantly. However, this could also lead to new artifacts. For instance, the tile like structure appearing on Fig. \ref{real} (d) is due to these sub-images. But as seen from this image, these artifacts are less significant to affect the outcome.

\section{Conclusions}
We have proposed a technique to improve the quality of the data in astronomical observations using subspace decomposition. Application of this technique to real data has shown us that it is superior to the normally used averaging techniques.

\section{Acknowledgement}
The author would like to thank Ger de Bruyn and Jan Noordam for posing this problem. This work
was supported by LOFAR and SNN. LOFAR is being funded by the European Union, European Regional Development Fund, and by ``Samenwerkingsverband Noord-Nederland'', EZ/KOMPAS.
\bibliographystyle{IEEE}

\end{document}